\documentclass[aps,prl,twocolumn]{revtex4}
\usepackage{epsfig}
\usepackage{color}   % colors
\usepackage{graphicx}

\begin{document}

\title{Brownian Motion of Graphene}
\author{O.M. Marag\`o$^{1}$, F. Bonaccorso$^{2}$, R. Saija$^{3}$, G.Privitera$^{2}$, P.G. Gucciardi$^{1}$, M.A. Iat\`{\i}$^{1}$,G. Calogero$^{1}$, P.H. Jones$^{4}$, F. Borghese$^{3}$, P. Denti$^{3}$,V. Nicolosi$^{5}$,A.C. Ferrari$^{2}$}\email{acf26@eng.cam.ac.uk}

\affiliation{$^1$ CNR-Istituto per i Processi Chimico-Fisici, I-98158 Messina, Italy}
\affiliation{$^2$ Engineering Department, University of Cambridge, Cambridge CB3 0FA,UK}
\affiliation{$^3$ Dipartimento di Fisica della Materia e Ingegneria Elettronica, Universit\'a di Messina, I-98166 Messina, Italy}
\affiliation{$^4$ Department of Physics and Astronomy, University College London, London WC1E 6BT, UK}
\affiliation{$^5$ Department of Materials, University of Oxford, Oxford OX1 3PH,UK}

\begin{abstract}
We study the Brownian motion (BM) of optically trapped graphene flakes. These orient orthogonal to the light polarization, due to the optical constants anisotropy. We
explain the flake dynamics, measure force and torque constants and derive a full electromagnetic theory of optical trapping. The understanding of two dimensional BM paves
the way to light-controlled manipulation and all-optical sorting of biological membranes and anisotropic macromolecules.
\end{abstract}
\maketitle

The random motion of microscopic particles in a fluid was first observed in the late eighteenth century, and goes by the name of Brownian motion(BM)\cite{Brown1828}. This was ascribed to thermal agitation\cite{Gouy1888}, leading to Einstein's predictions of the resulting particle displacements\cite{Einstein1905}. BM is ubiquitous throughout physics, chemistry, biology, and even finance. It can be harnessed to produce directed motion\cite{Jones2004}. It was also suggested that thermally activated BM may be responsible for the movement of molecular motors, such as myosin and kinesin\cite{Astumian97}. When a Brownian particle (BP), i.e. a particle undergoing BM, is subject to an external field, e.g. a confining potential, the fluid damps the BM and, in a high damping regime, such as for a BP in water, the confining potential acts as a cut-off to the BM dynamics. This is free for short times (high frequency limit), while is frozen at longer times (low frequency limit)\cite{Ornstein30}. These processes have perfect ground in experiments with optical traps, where a BP is held by a focused laser beam, i.e. an optical tweezers\cite{Ashkin86}. In this context, BM can be utilized to investigate the properties of the surrounding environment\cite{Svoboda94,Martin06}, as well as of the trapped particle, and for accurate calibration of the spring constants of the optical harmonic potential\cite{Pralle99main,Rohrbach05}.

Dimensionality plays a special role in nature. From phase transitions\cite{LeBellac}, to transport phenomena\cite{Heinzel}, two-dimensional (2d) systems often exhibit a
strikingly different behavior\cite{LeBellac}. Nanomaterials are an attractive target for optical trapping\cite{Peter06,Peter07,Marago08main}. This can lead to top-down
organization of composite nano-assemblies\cite{Peter06}, sub-wavelength imaging by the excitation and scanning of nano-optical probes\cite{Peter07}, photonic force
microscopy with increased space and force resolution\cite{Marago08main}. Graphene\cite{Novoselov04} is the prototype 2d material, and, as such, has unique mechanical,
thermal, electronic and optical properties\cite{Geimr}. Here, we use graphene as prototype material to unravel the consequences of BM in 2d.
\begin{figure}
\centerline{\includegraphics [width=100mm]{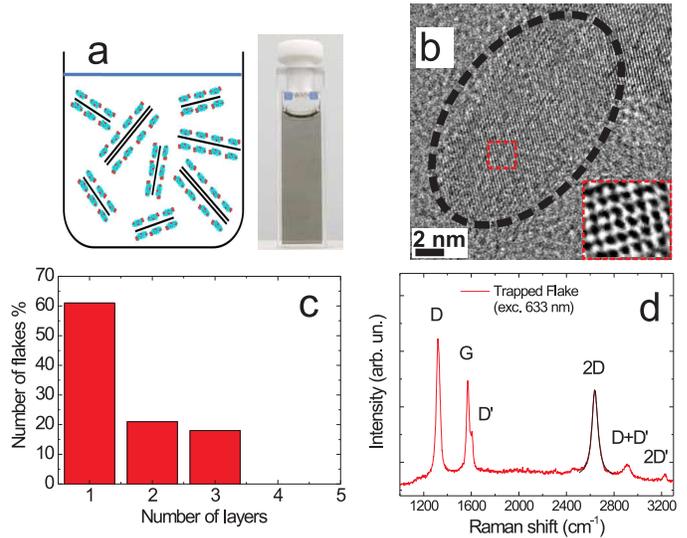}}\caption{a)Scheme of water dispersion of graphene. b)TEM of representative flake, showing the typical honeycomb
structure.c)Number of layers per flake from TEM images, showing up to 60$\%$ SLG.d), Raman spectrum of an optically trapped flake, for 633nm trapping and excitation
wavelength.}\label{Fig1}
\end{figure}

\begin{figure}
\centerline{\includegraphics [width=65mm]{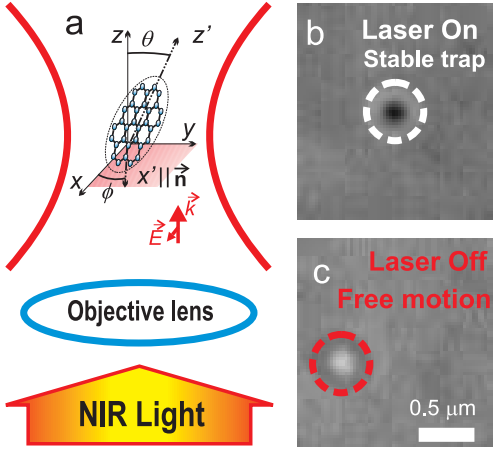}} \caption{a)Experimental setup. A laser beam is expanded to over-fill the back aperture of a high NA lens.
Geometry, relevant angles and axes are also shown. b) Laser is switched on and the flake is drawn into the optical trap. c) Laser is switched off and the flake released.}
\label{Fig2}
\end{figure}

Graphene is dispersed by processing graphite in a water-surfactant solution, Fig.\ref{Fig1}a. We do not use any functionalization nor oxidation, to retain the pristine electronic structure in the exfoliated monolayers\cite{Hernandez08main,Lotya09main,Sun10}. We use di-hydroxy sodium deoxycholate surfactant. High resolution
Transmission Electron Microscopy (HRTEM) shows$\sim$10-40nm flakes. Fig.\ref{Fig1}b is an image of one such flake, with the typical honeycomb lattice. By analyzing over a
hundred flakes, we find$\sim60\%$ single-layer (SLG), much higher than previous aqueous\cite{Lotya09main} and non-aqueous dispersions\cite{Hernandez08main}, and the
remainder bi-and tri-layers (Fig.\ref{Fig1}c).We then place 75$\mu$l in a chamber attached to a piezo-stage with 1nm resolution. Optical trapping is obtained by focusing
a near-infrared (NIR) (830nm) or a Helium-Neon (633nm) laser through a 100$\times $ oil immersion objective (NA$=1.3$) in an inverted configuration. The
latter is coupled to a spectrometer through an edge filter. This allows us to use the same laser light both for optical trapping and Raman scattering, realizing a Raman
optical tweezers (ROT) to directly probe the structure of the trapped flake, Fig.\ref{Fig1}d. In both setups, the particles are imaged through the same objective
(Fig.\ref{Fig2}a) that focuses the trapping light onto a CCD camera, with diffraction limited resolution. Fig.\ref{Fig2}c shows an flake drawn into the optical trap when
the laser is switched on. When the laser is switched off (Fig.\ref{Fig2}c) the flake is released and diffuses from the trap region. In both set-ups, the minimum trap
power is$\sim1-2$mW.

A typical Raman spectrum of trapped flakes measured at 633nm is plotted in Fig.\ref{Fig1}d. Besides the G and 2D peaks, this has significant D and D' intensities, and the combination mode D+D'$\sim$2950cm$^{-1}$. The G peak corresponds to the E$_{2g}$ phonon at the Brillouin zone centre. The D peak is due to the breathing modes of sp$^2$ rings and requires a defect for its activation by double resonance (DR)\cite{Ferrari07,Ferrari00,Tuinstra}. The 2D peak is the second order of the D peak. This is a single line in SLG, whereas it splits in four in bi-layer graphene, reflecting the evolution of the band structure\cite{Ferrari07}. The 2D peak is always seen, even when no D peak is present, since no defects are required for the activation of two phonons with the same momentum, one backscattering from the other. DR can also happen intra-valley, i.e. connecting two points belonging to the same cone around \textbf{K} or \textbf{K'}. This gives rise to the D' peak. The 2D' is the second order of the D' peak. The large D peak in Fig.\ref{Fig1}d is not due to a large amount of structural defects, otherwise it would be much broader, and G, D' would merge\cite{Ferrari00}. We assign it to the edges of our sub-micron flakes\cite{Casiraghi09}. We note that the 2D band, although broader than pristine graphene, is still fit by a lorentzian. Thus, even if the flakes are multi-layers, they are electronically almost decoupled\cite{Latil07}.

Positional and angular displacements of a flake in the optical trap are detected by back focal plane (BFP) interferometry using the forward scattered light from the
trapped particle\cite{Pralle99main,Marago08main}. The BFP interference pattern is determined by the flake orientation through the relevant angles ($\phi, \theta$). Since
the trapped flake is aligned with the $yz$ plane, fluctuations occur in the small angle limit, $\phi \ll 1, \theta \ll 1$, and the particle tracking signals in the
cartesian directions are: $ S_x \sim \beta_x \left( X - \textrm{a}\, \phi + \textrm{b} \, \theta \right);\,  S_y \sim \beta_y \left(Y + \textrm{c}\, \phi
\right);\, S_z \sim \beta_z  Z $, where $\beta_i$ are detector calibration factors, $X,Y,Z$ center-of-mass coordinates, a,b,c depend on flake
geometry and optical constants. $S_z$ is not affected by angular motion for small angles.

\begin{figure}
\centerline{\includegraphics [width=90mm]{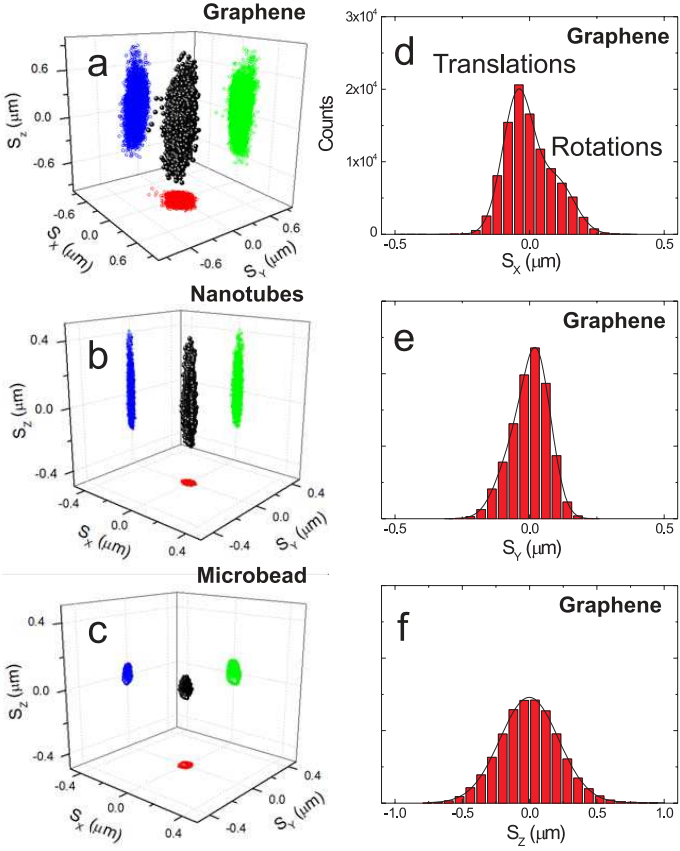}} \caption{a) Three-dimensional BM of a flake as compared to that of b) a nanotube bundle, and c) a latex
microbead. $10^4$ data-points are extracted from the $S_i(t)$. d,e) Histograms of the transverse signals $S_x(t),S_y(t)$. In the transverse direction, both translational
and angular fluctuations are superposed. The difference in the root mean square widths of the fluctuations in $x$ and $y$ arises from the flake shape, optical anisotropy,
and different curvatures of the optical potential in the directions parallel and perpendicular to the initial polarization. For each graph the QPD voltage-to-position
calibration factors $\beta_i$ are obtained using the calculated mobility coefficients and amplitude of the signals' autocorrelation functions $C_{ii}(0)=\beta_i^2 k_B
T/k_i$ for the position fluctuation contributions only. The root mean squares transverse displacements from a gaussian fit: $\sqrt{\langle x^2 \rangle}=57\pm 2$nm;
$\sqrt{\langle y^2 \rangle}=53\pm2$nm. For rotations: $\sqrt{\langle \phi^2 \rangle}=0.11\pm0.01$rad$\sim$7$\deg$, consistent with the small angle approximation.f)
Longitudinal signal $S_z(t)$, only due to center of mass fluctuations;$\sqrt{\langle z^2 \rangle}=217\pm5$nm} \label{Fig3}
\end{figure}

Fig.\ref{Fig3}a visualizes the BM of a graphene flake, compared with a nanotube bundle (Fig.\ref{Fig3}b) and a spherical latex microbead (Fig.\ref{Fig3}c), measured in the same experimental conditions. It is clear that these 2d, 1d and 3d objects exhibit distinct behaviors. The difference in the dynamics is due to the particle shape and optical properties. For a spherical particle the hydrodynamics is isotropic. Therefore, the different extent of fluctuations from equilibrium is only due to the anisotropy of the optical potential\cite{Rohrbach05,Borghese07main}. For a linear nanostructure anisotropic hydrodynamics leads to a much increased mobility along the optical axis\cite{Marago08main}. In contrast, the graphene flake has increased fluctuations in both longitudinal and transverse directions, that we ascribe to a higher contribution from rotational motion with respect to nanotubes. As discussed later, this is a fingerprint of the 2d geometry, yielding an increased sensitivity to angular fluctuations about the optical axis. The large optical anisotropy of graphene enhances this further, aligning the flake orthogonal to the light polarization. The effect of the rotational BM is illustrated in the histograms of Figs.\ref{Fig3}d,e,f, where the contribution from a superposition of translational and rotational fluctuations is seen in the transverse directions, but is absent in the longitudinal.

In order to extract quantitative data, we first analyze the flakes hydrodynamics, which encompasses translational and rotational motions. The viscous drag and torque are
described by the anisotropic mobility tensors\cite{Perrin34main,Han06} $\Gamma^t_{ij}$, for translations and $\Gamma^r_{ij}$, for rotations. These are related to the
fluid dynamical viscosity $\eta$ ($0.911$mPa~s for water at 24$^{\circ}$C) and particle size. We approximate graphene flakes as extremely flat ellipsoids, with transverse
size $\Delta$ much larger than their height $h$. This allows us to exploit the analytic solutions for uniaxial ellipsoids\cite{Perrin34main}. The
hydrodynamic mobilities are then only function of $\eta$ and $\Delta$: $\Gamma_{\|}\approx\frac{1}{8 \eta \Delta}, \Gamma_{\perp}\approx\frac{3}{16 \eta \Delta},
\Gamma^r\approx\frac{3}{4 \eta \Delta^3}$. The rotational mobility is the same for any axis through the center-of-mass, while the parallel
translation mobility is 2/3 the perpendicular one. From TEM, $\Delta\sim25$nm. Thus $\Gamma_{\|}\sim5.49\mu$m/(fN s), $\Gamma_{\perp}\sim8.26\mu$m/(fN s), $\Gamma^{r}\sim52.6$/(fN~nm~s).

We describe the Brownian dynamics of trapped graphene using uncoupled Langevin equations\cite{LeBellac}:
\begin{eqnarray}
\partial_{t} X_i(t)=- \omega_{i} X_i(t) + \xi_i(t),& i=x,y,z &
\label{Langevin1} \\
\partial_t \phi(t) = -\Omega_{\phi} \phi(t) + \xi_{\phi}(t), & &
\label{Langevin2}\\
\partial_t \theta(t) = -\Omega_{\theta} \theta(t) + \xi_{\theta}(t), & & \label{Langevin3}
\end{eqnarray}
where $X_i, \phi, \theta$ are stochastic variables associated with position and angular coordinates, $\xi_i(t)$ random noise sources with zero mean and variance $\langle\xi_i(t)\xi_i(t+\tau)\rangle=2k_B T\Gamma_i\delta(\tau)$, $\omega_{i}=\Gamma_i k_i$, $\Omega_{\phi}=\Gamma^r k_{\phi}$ and $\Omega_{\theta}=\Gamma^r k_{\theta}$ relaxation frequencies related to force and torque constants and mobility tensor components. The confining potential torque on the lab axes is only relevant for orientational dynamics, while not affecting center-of-mass motion in a small angle regime. Also, due to the strong yz-alignment, and because angular fluctuations are small, the radiation torque along $y$ (affecting $\theta$) is small and $\Omega_{\theta}\approx 0$.

We now evaluate the temporal correlations between the particle tracking signals, which yield the trap parameters\cite{Martin06}. For a non-spherical particle, correlation
function analysis reveals information about center-of-mass and angular fluctuations, hence on trap force and torque constants\cite{Marago08main}. For a strongly aligned
2d flake, the autocorrelation of the transverse tracking signal $C_{ii}(\tau) = \langle S_i(t)S_i(t + \tau)\rangle$ decays with lag time $\tau$ as a double
exponential corresponding to positional and angular relaxation frequencies $\omega_{i}$ ($i=x,y$), $\Omega_{\phi}$ , whereas, since the stochastic variables are
uncorrelated in the small angle regime, the cross-correlations $C_{xy}(\tau)=\langle S_x(t)S_y(t + \tau)\rangle$ of the transverse signals decay as a \textit{single}
exponential, with a relaxation rate corresponding to $\Omega_{\phi}$.

\begin{figure}[tbp]
\centerline{\includegraphics [width=95mm]{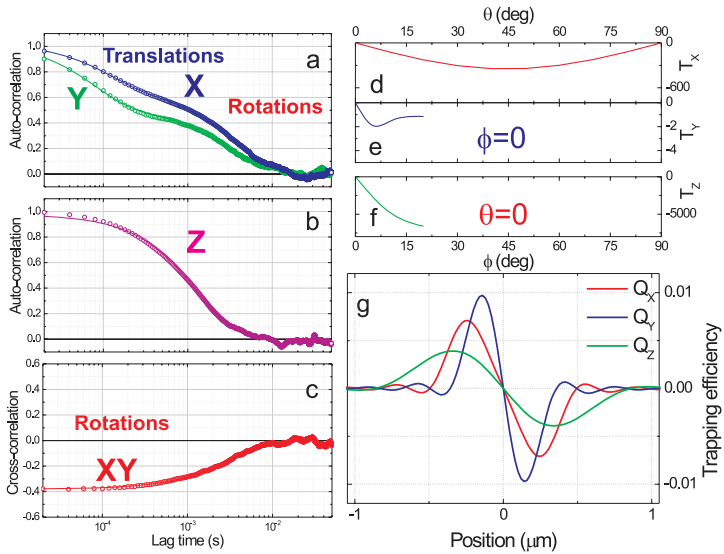}}\caption{a)Transverse (x,y), b) longitudinal (z) autocorrelations. The double exponential decay in a) is due to the different time scales of translational and angular BM. A single exponential is seen in the axial (z) direction. Solid lines are
exponential fits.c)Transverse signal cross-correlations revealing only angular fluctuations, having decay rate consistent with the slower part of the transverse
auto-correlation. The negative sign is related to the 2d geometry, resulting in an opposite x,y phase during rotation.d,e,f)Calculated optical torque components. Stable
orientation arises for a flake in the yz-plane (orthogonal to the polarization axis). For $T_y$ and $T_z$ no data is shown for $\phi$,$\theta>$20$^{\circ}$ because the
flake is expelled from the trap by radiation pressure. The polarization torque ($T_z$) is two orders of magnitude higher than the other components.g) The optical trapping
efficiency components $Q_i=c {\rm F}_{\rm Rad,i}/nP$ (c velocity of light, n=1.33 water refractive index and P laser power) are proportional to the optical force. For
small displacements the force follows Hooke's law and the derivatives at the equilibrium position define $k_x,k_y,k_z$} \label{Fig4}
\end{figure}

These allow us to derive the optical force constants from the relaxation frequency measurements. Fig.\ref{Fig4}a is a representative autocorrelation function analysis of the transverse tracking signals $C_{ii}(\tau) = \langle S_i(t)S_i(t +\tau)\rangle$ ($i = x,y$). These data are well fitted by two exponentials with $\omega_x=(8.6\pm 0.2)\times 10^3$s$^{-1}$, $\omega_y=(12.9\pm 0.3)\times 10^3$s$^{-1}$ for the translational decay rates and $\Omega_{\phi}=(3.0\pm 0.1)\times 10^2$s$^{-1}$ for the angular ones (obtained as average values from the $C_{xx}$ and $C_{yy}$ slow relaxation rate). Fig.\ref{Fig4}b shows that the autocorrelation of the axial signal $C_{zz}$ is well fitted by a single exponential with $\omega_z=(7.70\pm 0.05)\times 10^2$s$^{-1}$. In Fig.\ref{Fig4}c the cross-correlation of the transverse signals $C_{xy}$ is fitted by a single exponential with $\Omega_{\phi}=(2.90\pm 0.05)\times 10^2$s$^{-1}$, consistent with the value obtained from the autocorrelation functions. Repeating these measurements over ten different flakes, and using our estimation of the hydrodynamic mobility parameters, we obtain the spring constants $k_i=\omega_i/\Gamma_i$ to be $k_x = 1.1\pm 0.4$pN/$\mu$m, $k_y=1.3\pm 0.5$pN/$\mu$m, $k_z=0.08\pm0.03$pN/$\mu$m and torque constant about the propagation direction $k_{\phi}=\Omega_{\phi}/\Gamma^r=9\pm3$fmN$\cdot$nm/rad, where the uncertainty takes into account the 40$\%$ spread on flake size. Note how the measured force constants only depend on the flake transverse size $\Delta$, not on thickness $h$, because of the 2d geometry that strongly effects both hydrodynamics and radiation force and torque.

We calculate the radiation force $\mathbf{F}_{\rm Rad}$ and torque $\bf{\Gamma}_{\rm Rad}$ within the transition matrix approach\cite{Borghese07main,Rosalba08main}. We
first consider the incident field in the focal region of a high NA lens\cite{Borghese07main}. $\mathbf{F}_{\rm Rad}$ and $\bf{\Gamma}_{\rm Rad}$ are derived from the
linear and angular momentum conservation for field and graphene\cite{Rosalba08main}. The dielectric constant of graphene is highly
anisotropic\cite{Casiraghi07,Kravetsmain,leftacsnano} with components $\varepsilon_{\perp}$ and $\varepsilon_{\parallel}$ in the directions perpendicular and parallel to
the c axis (see Fig.\ref{Fig2}a)\cite{Kravetsmain}. At 830nm the graphene refractive index is $n_{\perp}=3+i\,1.5$; $n_{\parallel}=1.694$\cite{Kravetsmain}. Note that the
n$_{\perp}$ imaginary part yields a large absorption, while for n$_{\parallel}$ this is negligible. We then calculate $\mathbf{F}_{\rm Rad}(\mathbf{r})$, $\mathbf{r}$
being the flake center of mass relative to the focal point (Fig.\ref{Fig2}a). Trapping occurs when $\mathbf{F}_{\rm Rad}(\mathbf{r})$ vanishes with a negative derivative,
Fig.\ref{Fig4}g. Due to the flake and field symmetry in the focal region\cite{Borghese07main}, trapping occurs on the optical axis. For small displacements from
equilibrium, the trap is well approximated by an harmonic potential $V(x_{i})=\frac{1}{2}\sum_{i=x,y,z}k_{i}x_{i}^{2}$, with spring constants $k_{z}<k_{x},k_{y}$
depending on both flake geometry and gaussian beam power and polarization\cite{Borghese07main,Marago08main}.

The flake orientation is specified by the angles $\vartheta$, $\varphi$, Fig.\ref{Fig2}a. For each orientation we determine the center of mass trapping position, then the
torque relative to each axis at that position (Figs.\ref{Fig4}d,e,f): $\mathbf{T}=\frac{8\pi k}{n^2|E_0|^2\sigma_{\rm T}}\bf{\Gamma}_{\rm Rad}$, where $E_0$ is the
incident field amplitude, $\sigma_{\rm T}$ the flake extinction cross section, $n=1.33$ the water refractive index. Orientational stability occurs when ${\bf T}$ vanishes
with negative derivative with respect to both $\vartheta$, $\varphi$ (Figs.\ref{Fig4}d,e,f). We get that stable trapping is achieved when the flake plane is parallel to
$yz$. When the polarization axis lies on the flake plane (e.g. when the flake is parallel to $xy$ or $xz$), the radiation pressure is so strong that the flake is pushed
out of the trap. This is a consequence of the large imaginary part of $\varepsilon_{\perp}$. As shown in Figs.\ref{Fig4}d,e, the flake is stable under small
angle rotations around its equilibrium orientation, while for larger values of $\phi$ and $\theta$ ($>$20$^{\circ}$) it is expelled from the trap by radiation pressure.
Moreover, the polarization torque ($T_z$) is a hundred times larger than the other components because of optical constants anisotropy.

The graphene dimensionality and strong anisotropy in both optical and hydrodynamic properties determine the flake stability. The ability to discriminate linear and
angular motion is key to understand optical trapping of planar structures. Our results and methodology are generic, and can be extended to any 2d structure, such as
biological membranes. These contain micro-domains that play a role in signal transduction and trafficking\cite{Mukherjee04}. The diffusion and BM of planar structures in
a restricted, anisotropic environment could allow to understand the dynamics and function of these domains. The optical trap is an ideal environment for
spectroscopic and mechanical probing of such structures, linking their BM dynamics to their form, interactions and ultimate function.

We thank E. Lidorikis for useful discussions. We acknowledge funding from EPSRC grants EP/G042357/1, ERC NANOPOTS, Royal Society Brian Mercer Award for Innovation. F.B. acknowledges funding from a Newton International Fellowship. A.C.F. is a Royal Society Wolfson Research Merit Award holder.

\end{document}